\documentclass[fleqn,twoside]{article}
\usepackage{espcrc2}

\usepackage{graphicx}
\usepackage[figuresright]{rotating}

\newcommand{\AmS}{{\protect\the\textfont2
  A\kern-.1667em\lower.5ex\hbox{M}\kern-.125emS}}

\def\beq{\begin{equation}}
\def\eeq{\end{equation}}
\def\bea{\begin{eqnarray}}
\def\eea{\end{eqnarray}}
\def\bem{\begin{math}}
\def\eem{\end{math}}
\def\bit{\begin{itemize}}
\def\eit{\end{itemize}}
\def\bla{\begin{flushright}}
\def\ela{\end{flushright}}
\def\qq2{$Q^2$}               
\def\aa1{$A_1(x,Q^2)$}        
\def\ff1{$F_1(x,Q^2)$}        
\def\gg1{$g_1(x,Q^2)$}        
\newpage

\title{Some methods to evaluate complicated Feynman integrals
}

\author{A.V. Kotikov\address[MCSD]{ BLThPh,
Joint Institute for Nuclear Physics,
141980 Dubna, Russia}}

\begin{document}

\begin{abstract}
I discuss a progress in calculations of Feynman integrals 
based on the Gegenbauer Polynomial Technique and 
the Differential Equation Method. 
\vspace{-0.5cm}
\end{abstract}

\maketitle


\section{ Introduction }

Last years there was an essential progress in calculations
of Feynman integrals. It seems that most important results have
been obtained for two-loop four-point massless Feynman diagrams:
in on-shall case (see \cite{Smirnov,GeRede}) and for a class of off-shall
legs (see \cite{Offshall}). A review of the results can be found in
\cite{GeReconf}. Moreover, very recently results for a class of these
diagrams have been obtained
\cite{mass} in the case when some propagators have a nonzero mass.

In the letter, I review very shortly results obtained with help of 
two methods for calculations of Feynman diagrams (for details,
see \cite{Winter02}): 
the Gegenbauer Polynomial Technique \cite{2} (see also \cite{3,4}) and
 the Differential Equation Method (DEM) \cite{DEM1}.
The additional information about a modern progress in calculations
of Feynman integrals can be found, for example, also in recent articles 
\cite{Passarino}.

\section{Applications of the Gegenbauer Polynomial Technique}

The Gegenbauer Polynomial Technique
has been used for evaluation
of very complicated Feynman diagrams  (see also \cite{4})
which contribute mostly in calculations based on
various type of $1/N$ expansions: 
\begin{itemize}
\item
In the calculation (in \cite{QED3}) of the next-to-leading (NLO)
corrections to the value of dynamical mass generation (see \cite{Nash})
in the framework of three-dimensional Quantum Electrodynamics. 
\item
In the evaluation (in \cite{KoKo}) of the correct value of 
of the leading order contribution to the $\beta$-function
of the $\theta$-term in Chern-Simons theory. The $\beta$-function
is zero in the framework of usual perturbation theory but it takes nonzero
values in $1/N$ expansion (see \cite{korea}).
\item
In the evaluation (in \cite{Fadin}) of NLO corrections to the value of
gluon Regge trajectory (see discussions in \cite{Fadin} and references 
therein).
\item
In the calculation (in \cite{KoLi}) of the next-to-leading 
corrections to the BFKL intercept of spin-dependent part of high-energy
asymptotics of hadron-hadron cross-sections. 
\item
In the calculation (in \cite{KoLi,KoLi1}) of the next-to-leading 
corrections to the BFKL equation at arbitrary conformal spin.
\item
In the evaluation (in \cite{BrKo}) of the 
most complicated parts of ${\rm O}(1/N^3)$ contributions to 
critical exponents of $\phi^4$-theory,
for any spacetime dimensionality $D$.
\end{itemize}

\section{The recent progress in calculation of Feynman integrals by
the DEM.} 

\begin{itemize}
\item
The articles \cite{FKV1} and \cite{FKV}: 

{\bf a)} The set of two-point two-loop Feynman diagrams 
with one- and two-mass
thresholds has been evaluated by DEM (see Fig.1).
The results are presented  in Ref \cite{FKV1,FKV} and in the review 
\cite{Winter02}.
Some of them have been
known before (see 
\cite{FKV1}). The check of the results 
has been
done by Veretin programs (see discussions in 
\cite{FKV1} and references therein).
   \begin{figure}[tb]
\vskip -0.3cm
\includegraphics[width=1.8in,,angle=-90]{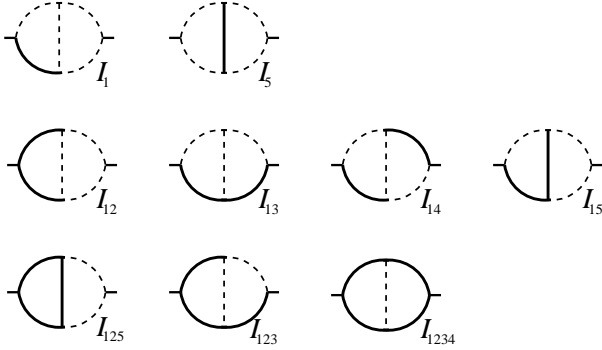} 
\vskip -0.7cm
 \caption{
Two-loop self-energy diagrams.
Solid lines denote propagators with the mass $m$; dashed lines
denote massless propagators.
\vskip -0.8cm}
 \end{figure}

{\bf b)} The set of three-point two-loop Feynman integrals
 with one- and two-mass
thresholds has been evaluated (the results of some of them has been
known before (see \cite{FKV1})) by a combination of DEM and Veretin programs
for calculation of first terms in small-moment expansion of Feynman diagrams
(see discussions in \cite{FKV1} and references therein).
\item
The article \cite{FKK}: 
The full set of two-point two-loop on-shell master diagrams 
has been evaluated by DEM. The check of the results has been
done by Kalmykov programs (see 
discussions in 
\cite{FKK,FKK1} and references therein). 
\item
The article \cite{GeRede}: 
The set of three-point and four-point two-loop massless Feynman diagrams 
has been evaluated.
\end{itemize}

\vspace{-0.2cm}

{\bf Acknowledgments.}~~Author
 would like to express his sincerely thanks to the Organizing
  Committee of ACAT 2002
for the kind invitation.
He was supported in part by 
INTAS  grant N366.

\end{document}